\documentclass[twocolumn]{aastex63}
\usepackage{apjfonts}
\received{2020 October 18}
\revised{2020 December 20}
\accepted{2020 December 24}
\shorttitle{Deceleration of Blast Waves of Tycho's Supernova Remnant}
\shortauthors{Tanaka et al.}

\begin{document}

\title{Rapid Deceleration of Blast Waves Witnessed in Tycho's Supernova Remnant}

\correspondingauthor{Takaaki Tanaka}
\email{ttanaka@cr.scphys.kyoto-u.ac.jp}

\author[0000-0002-4383-0368]{Takaaki Tanaka}
\affiliation{Department of Physics, Kyoto University, Kitashirakawa Oiwake-cho, Sakyo, Kyoto 606-8502, Japan}

\author{Tomoyuki Okuno}
\affiliation{Department of Physics, Kyoto University, Kitashirakawa Oiwake-cho, Sakyo, Kyoto 606-8502, Japan}

\author[0000-0003-1518-2188]{Hiroyuki Uchida}
\affiliation{Department of Physics, Kyoto University, Kitashirakawa Oiwake-cho, Sakyo, Kyoto 606-8502, Japan}

\author[0000-0002-5092-6085]{Hiroya Yamaguchi}
\affiliation{Institute of Space and Astronautical Science, JAXA, 3-1-1 Yoshinodai, Sagamihara, Chuo, Kanagawa 252-5210, Japan}
\affiliation{Department of Physics, The University of Tokyo, 7-3-1 Hongo, Bunkyo, Tokyo 113-0033, Japan}

\author[0000-0002-2899-4241]{Shiu-Hang Lee}
\affiliation{Department of Astronomy, Kyoto University, Kitashirakawa Oiwake-cho, Sakyo, Kyoto 606-8502, Japan}

\author[0000-0003-2611-7269]{Keiichi Maeda}
\affiliation{Department of Astronomy, Kyoto University, Kitashirakawa Oiwake-cho, Sakyo, Kyoto 606-8502, Japan}

\author[0000-0003-2063-381X]{Brian J. Williams}
\affiliation{NASA Goddard Space Flight Center, Code 662, Greenbelt, MD 20771, USA}
%



\begin{abstract}
In spite of their importance as standard candles in cosmology and as major major sites of nucleosynthesis in the Universe, 
what kinds of progenitor systems lead to type Ia supernovae (SN) remains a subject of considerable debate in the literature. 
This is true even for the case of Tycho's SN exploded in 1572 although it has been deeply studied both observationally and theoretically. 
Analyzing X-ray data of Tycho's supernova remnant (SNR) obtained with Chandra in 2003, 2007, 2009, and 2015, we discover that the expansion before 2007 was substantially faster than radio measurements reported in the past decades and then rapidly decelerated during the last $\sim 15$ years. 
The result is well explained if the shock waves recently hit a wall of dense gas surrounding the SNR. 
Such a gas structure is in fact expected in the so-called single-degenerate scenario, in which the progenitor is 
a binary system consisting of a white dwarf and a stellar companion,  
whereas it is not generally predicted by a competing scenario, the double-degenerate scenario, which has a binary of two white dwarfs as the progenitor. 
Our result thus favors the former scenario. 
This work also demonstrates a novel technique to probe gas environments surrounding SNRs and thus disentangle the two progenitor scenarios for Type Ia SNe. 
\end{abstract}



\section{Introduction} \label{sec:intro}
Type Ia supernovae (SNe), in which white dwarfs in binary systems explode, are crucial in discussing chemical evolution of the Universe as major sites of nucleosynthesis \citep[e.g.,][]{thielemann1986}. 
They are important also as standard candles to measure distances in the field of cosmology \citep{riess1998,perlmutter1999}. 
For their progenitors, single-degenerate (SD) systems consisting of a white dwarf and a stellar companion have been considered \citep[e.g.,][]{whelan1973}. 
A competing scenario is that of double-degenerate (DD) systems, or binaries of two white dwarfs \citep[e.g.,][]{webbink1984}.  
However, even for the well-studied case of Tycho Brahe's SN of 1572,   
regarded as a ``standard'' Type Ia SN \citep{badenes2006,rest2008,krause2008}, it is still unclear which model is correct. 

Tycho's SN occurred in 1572 in the Milky Way Galaxy toward the direction of the constellation of Cassiopeia, 
and was one of the few SN explosions observable with naked eyes in historical records. 
Tycho Brahe observed it and left records of his observations in his book entitled ``De nova et nullius aevi memoria prius visa stella.'' 
 \cite{baade1945} derived the lightcurve from Tycho Brahe's record, and found that it is consistent with that of a Type Ia SN. 
In modern times, a scattered-light echo of the SN was discovered from a nearby dust cloud by \cite{rest2008}. 
The spectrum of the original SN emission was obtained from the echo spectrum, and was found to be consistent 
with a normal Type Ia SN spectrum (and inconsistent with sub- and over-luminous SNe Ia) \citep{krause2008}. 
This supports an earlier claim of a Type Ia origin based on X-ray spectroscopy of the emission from the ejecta from the remnant by \cite{badenes2006}. 

The progenitor system of Tycho's SN has actively been discussed in the literature and is still under considerable debate. 
In the case of the SD progenitor scenario, it is expected that the stellar companion should have survived the SN explosion. 
Therefore, such a star, if detected, can provide us with a decisive answer about whether the progenitor was an SD system or 
a DD system. 
Although a star named Star~G, which is located close to the center of the remnant, was claimed as a possible candidate for the ex-companion by \cite{ruiz-lapuente2004}, 
\cite{kerzendorf2009} and \cite{xue2015} disputed the claim. 
The situation is similar for the other proposed candidates, Stars~B and E \citep{ihara2007,kerzendorf2013,kerzendorf2018,ruiz-lapuente2019}. 
It seems that another method is needed to disentangle the two scenarios about the progenitor of Tycho's SN. In the SD scenario, mass accretion from a stellar companion onto a white dwarf leads to an explosion. \cite{hachisu1996} predicted that a mass-accreting white dwarf drives a strong wind from its surface, resulting in gas evacuation and formation of a low-density cavity surrounding the explosion site \citep{badenes2007,williams2011}. DD progenitors, on the other hand, are not expected to create such a gas structure. We hence aim to search for a wind cavity surrounding Tycho's supernova remnant (SNR) based on the measurement of the velocity of its expanding blast waves with multi-epoch Chandra X-ray data.

\section{Observations and Data Reduction} \label{sec:obs}
We analyzed data from observations of Tycho's SNR performed with ACIS-I on-board the Chandra X-ray Observatory in the years 2003, 2007, 2009, and 2015. 
The information on the observations is summarized in Table~\ref{tab:obs}. 
We reduced the data with the analysis software package CIAO version 4.11 with the calibration 
data base version 4.8.3. 
We reprocessed the datasets using the {\tt chandra\_repro} script in CIAO. 
For accurate expansion measurements, 
we aligned coordinates of each dataset to that of the dataset with ObsID = 10095, which has the longest exposure time. 
The alignment was performed by crossmatching point sources detected in the field-of-views 
using the {\tt wavdetect} tool in CIAO. 
Four datasets (ObsID = 8551, 10903, 10904, and 10906) were discarded here since {\tt wavdetect} detected too few point sources to proceed to further steps. 
Transformation matrices were computed with {\tt wcs\_match} by crossmatching detected point sources, and 
the coordinate values were reassigned with {\tt wcs\_update}. 
The matrices for most of the datasets describe translation, rotation, and scaling. 
For the dataset with ObsID = 10902, only translation is applied because of a small number of point sources detected. 
In Table~\ref{tab:obs}, we list the number of point sources used for the coordinate alignment and the parameters of the matrices for each dataset. 
We combined the six datasets from 2009 in the data analysis described below. 

\begin{deluxetable*}{cccccccc}[bth]
\tablenum{1}
\tablecaption{\label{tab:obs}Observation Log and Summary of Coordinate Alignment}
\tablewidth{0pt}
\tablehead{
\colhead{ObsID} & \colhead{Obs. Date\tablenotemark{a}} & \colhead{$T_{\rm exp}$ (ks)\tablenotemark{b}} & \colhead{$N_{\rm ps}$\tablenotemark{c}} & \colhead{$\Delta x$\tablenotemark{d}}& \colhead{$\Delta y$\tablenotemark{e}} & \colhead{$\theta$\tablenotemark{f}} & \colhead{$R_{\rm scale}$\tablenotemark{g}}
}
\startdata
3837 & 2003 Apr 29 & 146 & 11 & $-0\farcs1263$ & $0\farcs422666$ & $0\farcm11652$ & 1.000261  \\
7639 & 2007 Apr 23 & 109 & 8 & $-0\farcs351728$  & $-0\farcs472105$  & $-0\farcm67494$  & 1.002537 \\
8551\tablenotemark{i} & 2007 Apr 26 & 33 & \nodata & \nodata & \nodata & \nodata & \nodata \\
10093 & 2009 Apr 13 & 118 & 12 & $0\farcs179763$  & $0\farcs364343$  & $1\farcm16118 $ & 0.999776  \\
10094 & 2009 Apr 18 & 90 & 7 & $-0\farcs415714$  & $0\farcs0427164$  & $-0\farcm02778$ & 1.001382 \\
10095\tablenotemark{h} & 2009 Apr 23 & 173 & \nodata & \nodata & \nodata & \nodata & \nodata  \\
10096 & 2009 Apr 27 & 106 & 10 & $-0\farcs243941$  & $0\farcs168533$  & $1\farcm56732$ & 1.000538  \\
10097 & 2009 Apr 11 & 107 & 11 & $-0\farcs610284$  & $0\farcs18231$  & $1\farcm25904$ & 1.001266  \\
10902\tablenotemark{j} & 2009 Apr 15 & 40  & 4 & $0\farcs222649$  & $0\farcs344939$  & 0 (fixed)  & 1.0 (fixed)  \\
10903\tablenotemark{i} & 2009 Apr 17 & 24  & \nodata & \nodata & \nodata & \nodata  & \nodata   \\
10904\tablenotemark{i} & 2009 Apr 13 & 35  & \nodata & \nodata & \nodata & \nodata & \nodata \\
10906\tablenotemark{i} & 2009 May 03 & 41 & \nodata & \nodata & \nodata & \nodata & \nodata\\
15998 & 2015 Apr 22 & 147 & 14 & $-0\farcs298268$  & $0\farcs30051$  & $3\farcm59052$ & 1.000891  \\
\enddata
\tablenotetext{a}{Observation start date.}
\tablenotetext{b}{Exposure time.}
\tablenotetext{c}{Number of point sources used for coordinate alignment after deleting those with poor matches.}
\tablenotetext{d}{Shift along the $x$-axis of the image.}
\tablenotetext{e}{Shift along the $y$-axis of the image.}
\tablenotetext{f}{Rotation angle}
\tablenotetext{g}{Scale factor.}
\tablenotetext{h}{Used as a reference for coordinated alignment.}
\tablenotetext{i}{Not used in the analysis because of too few point sources detected for coordinated alignment.}
\tablenotetext{j}{ Only translation is applied in coordinate alignment.}
\end{deluxetable*}

\begin{figure*}[bth]
\epsscale{1.1}
\plotone{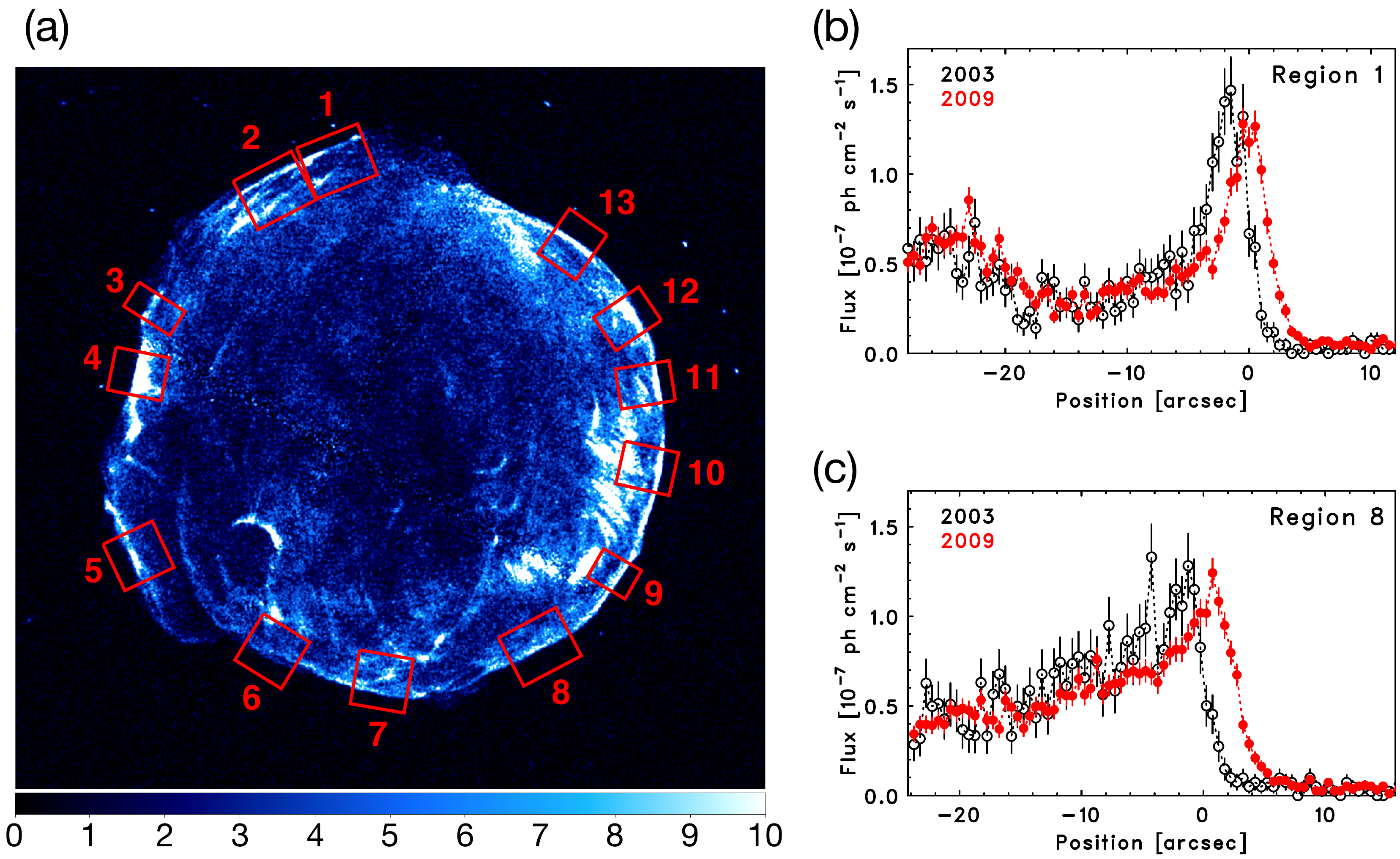}
\caption{
\label{fig:image}
(a) Chandra ACIS 
image of Tycho's SNR in the energy band of 4.1--6.1~keV obtained from the observations performed in 2009. 
North is up, and east is to the left. 
The color scale indicates flux from each pixel in units of $10^{-9}~{\rm photon}~{\rm cm}^{-2}~{\rm s}^{-1}$ 
with a pixel size of $0\farcs492 \times 0\farcs492$. 
The red rectangles are the regions used for the expansion measurements.  (b) Radial profiles observed 
in Region 1. 
The black and red points are from observations in 2003 and 2009, respectively. 
The origin of the horizontal axis corresponds to the location of the shock front in 2003. 
(c) Same as (b) but for Region 8. 
}
\end{figure*}

\section{Analysis and Results} \label{sec:ana}
Figure~\ref{fig:image}(a) shows a hard-band (4.1--6.1~keV) image of Tycho's SNR from the 2009 epoch. 
We chose the energy band for our expansion measurements since it is dominated by synchrotron radiation and clearly traces the blast wave of the SNR \citep{warren2005}. 
We picked up 13 regions with clear filamentary rims, and obtained profiles at each epoch to see the expansion of the blast waves. 
Examples from Regions 1 and 8 are presented in Figure~\ref{fig:image}(b) and (c), where expansion is clearly visible as already found by by \cite{katsuda2010} and \cite{williams2016}. 

Comparing the profiles in each region from two different epochs, we quantitatively measured the velocity and its evolution over the period from 
2003 to 2015. 
We applied here the same method as \cite{tanaka2020}. 
We histogrammed radial profiles so that each bin has a width of $0.5~{\rm arcsec}$. 
To measure the expansion between two epochs, we artificially shifted the profile obtained in one of the epochs and 
searched for the shift that gives the best match between the two profiles. 
We allowed shifts which are non-integer multiples of the bin width. 
Each time after shifting the histogram of a radial profile, we rebinned the shifted profile to a histogram with the same binning as 
the original histogram under an assumption that data distribute uniformly in each bin of the original histogram. 
The degree of matching between the two profiles was quantified with chi-squared defined as
\begin{eqnarray}
\chi^2 =  \sum_i \frac{(f_i - g_i)^2}{(df_i)^2 + (dg_i)^2}, 
\end{eqnarray}
where $i$ is the index for histogram bins, $f_i$ and $g_i$ are values of the bin $i$, and $df_i$ and $dg_i$ 
are their standard deviations. 
We fitted resulting $\chi^2$ profiles near their minimums with a quadratic function and obtained velocities that give the minimum $\chi^2$ ($\chi^2_\mathrm{min}$) 
and thus the best matches between the two radial profiles. 
We list $\chi^2_\mathrm{min}$ as well as degrees of freedom from each crossmatching in Table~\ref{tab:stat}. 
Parameter ranges that give $\chi^2 \leq \chi^2_\mathrm{min}+1$ are defined as the $1\sigma$ confidence intervals.

\begin{figure*}[bth]
\epsscale{0.8}
\plotone{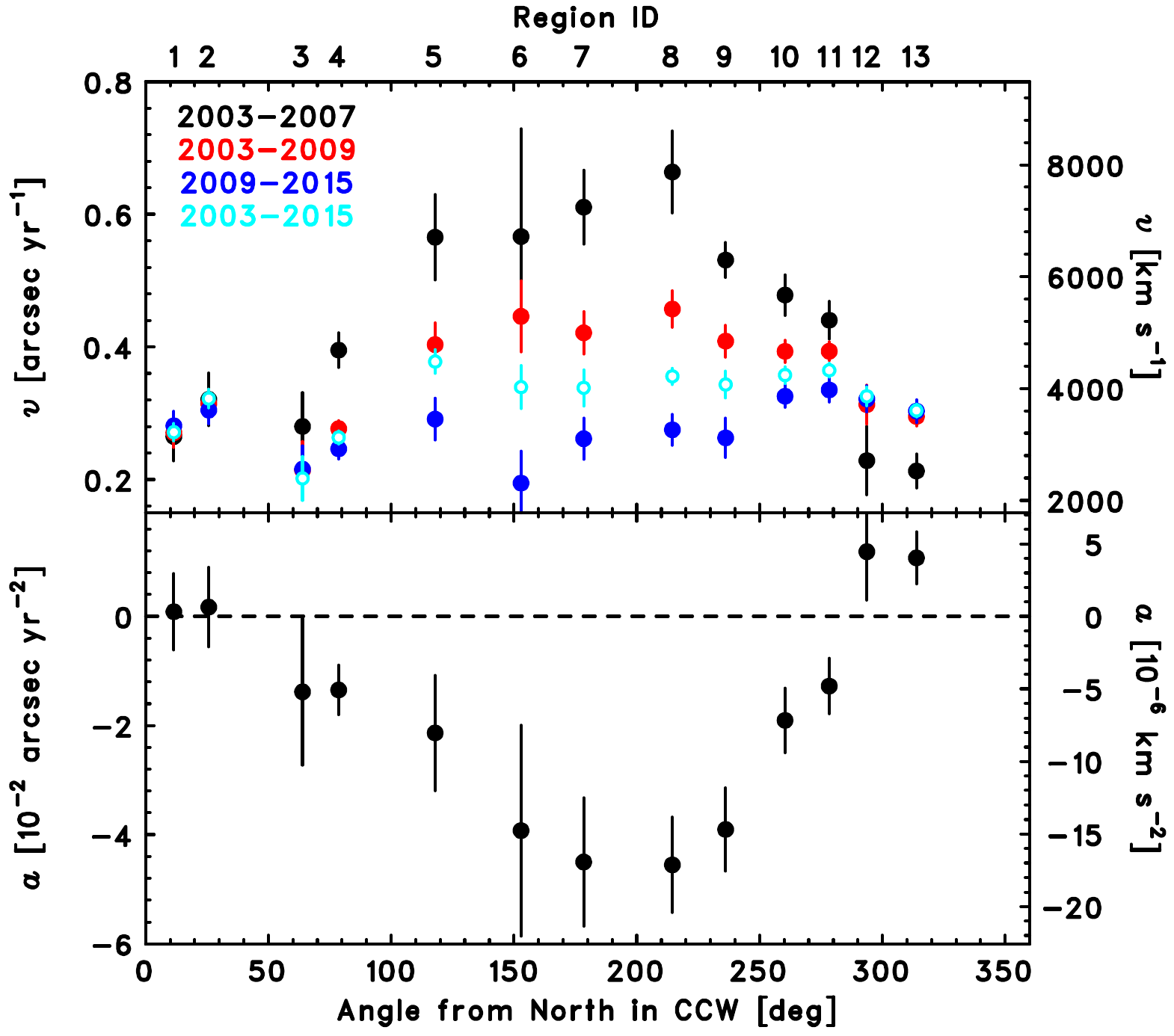}
\caption{
\label{fig:velocity} Velocity and acceleration of expansion of Tycho's SNR. The horizontal axis shows azimuthal angle 
in degrees from the north in counterclockwise direction. The numbers on top are corresponding region identifiers defined in Figure~\ref{fig:image}(a). 
The top panel shows velocities measured for the intervals of 2003--2007 (black filled circles), 2003--2009 (red filled circles), and 
2009--2015 (blue filled circles). 
For direct comparison with the result by \cite{williams2016}, velocities from the interval of 2003--2015 are also plotted (cyan small open circles). 
Plotted in the bottom panel are accelerations estimated under an assumption that they are constant 
over the time period between 2003 and 2015. 
The distance to Tycho's SNR is assumed to be 2.5~kpc in converting the angles to physical lengths.
}
\end{figure*}

Figure~\ref{fig:velocity} displays the results of the measurements for three different time intervals: 2003--2007, 2003--2009, and 2009--2015. 
In the interval of 2003--2007, the expansion velocity is relatively higher in the southern and western directions as pointed out by \cite{williams2016}.
Comparison between the data in 2003 and those in 2009 results in systematically lower velocities in those directions. 
The 2009--2015 interval gives even lower expansion velocities of the southern and western rims. 
Our measurements have revealed that the blast waves of Tycho's SNR are drastically being decelerated during the last $\sim 15~{\rm yr}$. 
The significance of the deceleration was quantified by fitting the velocities as a function of time with a constant function. 
A constant function as a null hypothesis was rejected with significances of $> 3 \sigma$ for Regions 4--10 as presented in Table~\ref{tab:stat}, 
indicating significant deceleration in these regions. 
To our knowledge, this is the first detection of on-going deceleration of blast waves of SNRs. 
We note that the 2003--2007 and 2003--2009 velocities are not independent since they were both derived using the same 2003 data. 
Likewise, the 2003--2009 and 2009--2015 measurements are not independent for the same reason. 
We, however, emphasize that the 2003--2007 and 2009--2015 measurements do not use the same dataset in common, and still indicate 
rapid deceleration.


\begin{deluxetable*}{cccccc}[ptbh]
\tablenum{2}
\tablecaption{\label{tab:stat}Statistics from Expansion Measurements}
\tablewidth{0pt}
\tablehead{
\colhead{Region ID} & \multicolumn{4}{c}{$\chi^2_\mathrm{min}$ (degrees of freedom)} &  \colhead{significance\tablenotemark{a}} \\
\colhead{} & \colhead{2003--2007} &  \colhead{2003--2009} &  \colhead{2009--2015} &  \colhead{2003--2015} &  \colhead{($\sigma$)}
}
\startdata
1  & 18.0 (22) & 20.4 (22) & 18.9 (22) & 17.0 (22) & 0.1\\
2  & 11.3 (20) & 14.0 (20) & 18.5 (20) & 13.5 (20) & 0.1\\
3  & 33.7 (35) & 76.5 (35) & 51.7 (35) & 47.9 (35) & 0.6\\
4  & 68.3 (47) & 68.1 (47) & 73.2 (47) & 71.6 (47) & 4.6\\
5  & 38.3 (40) & 45.7 (40) & 87.6 (40) & 52.2 (40) & 3.7\\
6  & 48.3 (40) & 58.3 (40) & 62.5 (40) & 47.1 (40) & 3.4\\
7  & 38.2 (40) & 41.1 (40) & 46.5 (40) & 45.9 (40) & 5.4\\
8  & 42.1 (40) & 36.0 (40) & 42.8 (40) & 32.0 (40) & 6.7\\ 
9  & 45.4 (25) & 45.6 (25) & 31.9 (25) & 44.7 (25) & 6.4\\
10 & 13.5 (15) & 14.6 (15) & 13.5 (15) & 15.5 (15) & 4.2\\
11 & 20.1 (15) & 23.2 (15) & 12.6 (15) & 19.7 (15) & 2.9\\
12 & 29.6 (40) & 34.7 (40) & 65.3 (40) & 61.9 (40) & 1.2\\
13 & 25.5 (27) & 29.7 (27) & 40.6 (27) & 34.5 (27) & 2.6\\
\enddata
\tablenotetext{a}{Significance with which a hypothesis of constant velocity between 2003 and 2015 is rejected.}
\end{deluxetable*}

\section{Discussion} \label{sec:dis}
We discovered rapid deceleration of the blast waves of Tycho's SNR by analyzing X-ray data obtained with Chandra in 2003--2015. 
Similar expansion measurements were performed by \cite{williams2016}, who obtained shock velocities based on comparisons 
of Chandra images in 2000/2003 and 2015. 
As a sanity check, we plotted the velocities for the time interval of 2003--2015 in Figure~\ref{fig:velocity} for direct comparison with the result by \cite{williams2016}.  
We found that the two measurements agree well with each other. 
We furthermore confirmed the rapid deceleration even when we applied the analysis method by \cite{williams2016}. 

Expansion of Tycho's SNR has also been measured in the radio band by several authors such as \cite{strom1982}, \cite{reynoso1997}, 
and \cite{williams2016}. 
The radio results tend to point to systematically lower velocities than our X-ray results, and do not show expansion velocities as high as what we observed before 2007. 
According to the results by \cite{reynoso1997}, the shock velocities measured between 1983/1984 and 1994/1995 are $0\farcs38~{\rm yr}^{-1}$ 
and $0\farcs35~{\rm yr}^{-1}$ in the locations corresponding to Regions~5 and 8, respectively. 
These values are closer to our 2003--2009 results than the 2003--2007 values. 
However, even without taking into account the possible systematic effects between the radio and X-ray measurements, 
our 2009--2015 velocities are still lower than the radio velocities and thus are consistent with recent deceleration of the blast waves. 

The systematics between the radio and X-ray results could be arising from observational or analysis methods, 
but could also be ascribed to physics related to the radio and X-ray emissions. 
Electrons emitting radio synchrotron photons have a much longer loss timescale than those emitting synchrotron X-rays. 
Thus, radio-emitting electrons would accumulate in the downstream region, which creates a plateau-like profile there \citep{Slane2014}.  
As the shock expands, the number of electrons accumulated in the downstream region would be increased, and the location of the radio emission peak would 
gradually be shifted toward downstream. 
On the other hand, X-rays would be emitted right at the shock because of much shorter lives of the radiating electrons. 
Such an effect may be able to account for the differences between radio and X-ray results.

It is notable that, after the deceleration, the expansion velocities reached $\sim 0\farcs3~{\rm yr}^{-1}$ regardless of direction (Figure~\ref{fig:velocity}). 
This can be interpreted as a consequence of the blast wave hitting a dense gas wall surrounding the SNR with approximately uniform density, i.e., a cavity wall structure. 
We performed one-dimensional spherically symmetric hydrodynamical simulations to test this assumption using the Virginia Hydrodynamics 1 (VH1) code \citep{BE2001}. 
Following a previous successful multi-wavelength model for this SNR by \cite{Slane2014}, 
we initiated the simulations using an SN ejecta with a mass of $1.4~M_\odot$, an explosion energy of $10^{51}~{\rm erg}$, 
and an exponential density profile. 
The ejecta first expands into a cavity with a uniform number density $n_{\rm b} = 0.3~{\rm cm}^{-3}$. 
The deceleration of the blast wave begins as it hits a dense gas cloud which is modeled as a density jump at a radius $R_{\rm c}$ 
from the explosion center with a spatial density gradient $dn_{\rm c}/dr$. 
We assume a uniform density $n_{\rm c} = 100~{\rm cm}^{-3}$ inside the wall, but to which 
our results are not sensitive since the blast waves at all regions experiencing a deceleration are found to be still climbing up the density gradient currently.

\begin{figure*}[hp]
\epsscale{0.8}
\plotone{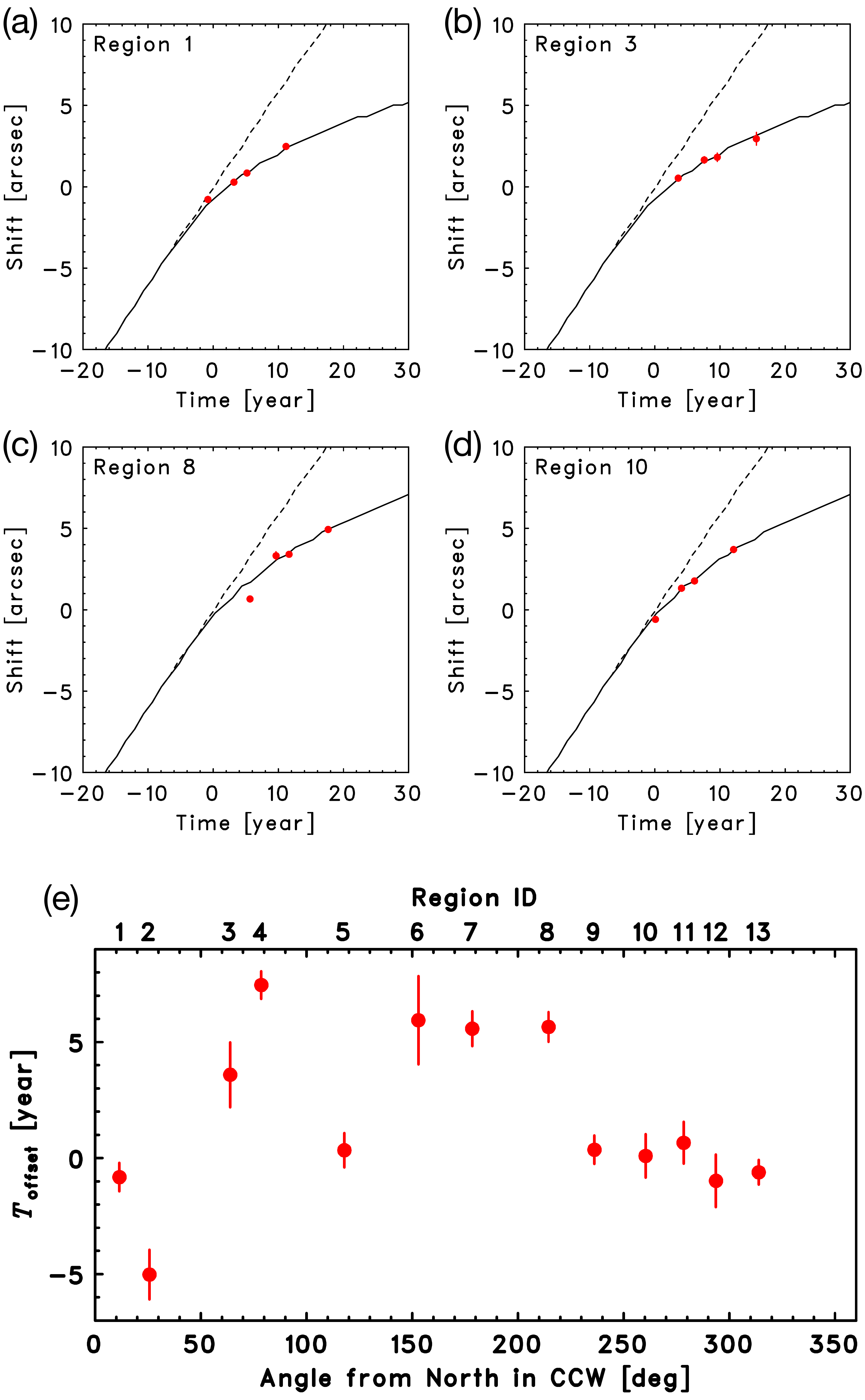}
\caption{
\label{fig:sim}  
(a) Expansion observed in Region 1 (red circles) as compared with the simulation result (black solid curve). 
Also shown are the result of the simulation without any shock--wall collisions (black dashed curve).  
(b)--(d) Same as (a) but for Regions 3, 8, and 10.  
(e) Relative timing of the shock--wall collision ($T_{\rm offset}$) obtained by fitting the data with the simulation results. 
The parameter $T_{\rm offset}$ is defined so that $T_{\rm offset} = 0$ corresponds to a case in which the blast wave started to interact with the wall in 2003. 
A larger $T_{\rm offset}$ means that the blast wave reached the wall at later times. 
}
\end{figure*}

We present the simulation result compared with the observational data in Figure~\ref{fig:sim}(a)--(d). 
Both $R_{\rm c}$ and $dn_{\rm c}/dr$ at the interface between the cavity and the wall are treated as model parameters 
to fit the observational data.    
We tried the five cases, $dn_{\rm c}/dr = 0$, $30$, $100$, $300$, and $1000~{\rm cm}^{-3}~{\rm pc}^{-1}$, 
for each region, and found that the data from Regions 1--4, 5--9, and 11--13 are best fitted with the $dn_{\rm c}/dr = 1000$, $300$, 
and $100~{\rm cm}^{-3}~{\rm pc}^{-1}$, respectively. 
In each region, we interpret that the forward shock reached the dense cloud at different ages due to 
a variation of $R_{\rm c}$ as a function of the azimuthal angle around the rim of the SNR. 
Fitting the model to the data, we determined when the blast wave started to interact with the wall in each region and present the result in Figure~\ref{fig:sim}(e). 
We found that larger decelerations, for example, in Regions 6, 7, and 8 can be nicely explained if the blast wave hit the wall at later times than 
locations with smaller decelerations such as Regions 1, 2, 12, and 13. 
This may suggest a somewhat asymmetric shape of the cavity wall.    
According to the simulation results, the expansion velocity is expected to stay almost constant well after the collision with the wall (Figure~\ref{fig:sim}(a)--(d)). 
Future observations of the SNR should be able to confirm this prediction.

If Tycho's SNR is surrounded by a cavity wall as our result implies, 
its dense gas would be detectable with, for example, CO line emission. 
A molecular bubble was indeed discovered by \cite{zhou2016} in a mapping observation of line emission at $230.5~{\rm GHz}$ from the $J = 2\textrm{--}1$ transition of 
$^{12}{\rm CO}$ molecules. 
The authors claimed that the blast waves of the SNR just reached the cavity wall based on the intensity ratios of the $^{12}{\rm CO}~(J=2\textrm{--}1)$ line 
to the $^{12}{\rm CO}~(J=1\textrm{--}0)$ and  $^{13}{\rm CO}~(J=1\textrm{--}0)$ lines, which is consistent with the implication of the rapid deceleration of the expansion 
that Chandra detected. 
Based on the radio data, the authors also estimated that the gas densities of the cavity is $\sim 0.02\textrm{--}0.1~{\rm cm}^{-3}$. 
If the molecular gas has a density of $\sim 10^{2-3}~{\rm cm}^{-3}$, a typical value for molecular gas, the density ratio becomes roughly consistent with 
the value that we assumed in the hydrodynamical simulations. 
Thus, it seems that the X-ray and radio data ``see'' the same gas structure surrounding the SNR. 
\cite{arias2019} recently detected low-frequency radio absorption in some regions of Tycho's SNR with the LOw Frequency ARray (LOFAR), 
indicating that the SNR is surrounded by an ionized thin cavity. 
The LOFAR result may imply that the shock has not yet reached the neutral molecular cloud but is running inside ionized gas 
present close to the edge of the cloud. 
 
The low-density cavity around Tycho's SNR is consistent with density estimates derived from previous expansion measurements 
in X-rays by \cite{katsuda2010} ($\lesssim 0.2~{\rm cm}^{-3}$), 
upper limit to thermal X-ray emission from shocked ambient medium by \cite{cassam2007}  ($< 0.6~{\rm cm}^{-3}$), and 
infrared flux measurements by \cite{williams2013} ($\sim 0.1\textrm{--}0.2~{\rm cm}^{-3}$). 
However, it is inconsistent with the density estimated from the degree of ionization of heavy elements in the shocked ejecta \citep{badenes2006} 
($\sim 1~{\rm cm}^{-3}$). 
A lower ambient density makes ejecta ionization slower and the ionization degree lower than what is observed now. 
One of the possible solutions to reconcile the contradiction would be non-uniform gas density in the bubble, e.g., denser gas near 
the progenitor system \citep{chiotellis2013,yamaguchi2014}.

It is rather unlikely that natal interstellar gas happens to be structured like a cavity wall surrounding Tycho's SNR, but, instead, 
it would be more natural to suppose that the gas structure is an imprint of activities of the progenitor system before the SN explosion. 
A detection of such circumstellar materials has been regarded as a smoking gun for an SD progenitor of a type Ia SN. 
The circumstellar materials strongly indicate that the progenitor should be a young system, making the DD scenario unlikely. 
In the SD scenario, 
a strong wind from the progenitor white dwarf during mass accretion \citep{hachisu1996} is one of the plausible mechanisms responsible for the 
formation of the cavity. 
Under the assumption that the cavity expands at the same velocity as the wind and 
that the wind velocity is constant, 
the radius of the cavity ($R_{\rm c} \sim 3~{\rm pc}$) can be explained if the wind duration is 
$t_{\rm w} = 3 \times 10^4 \, (v_{\rm w}/100~{\rm km}~{\rm s}^{-1})^{-1}~{\rm yr}$, where $v_{\rm w}$ is the wind velocity. 
Then the mass loss rate of the progenitor is estimated to be 
$\dot{M} = 9 \times 10^{-6} \, (n_{\rm b}/0.1~{\rm cm}^{-3})(v_{\rm w}/100~{\rm km}~{\rm s}^{-1})~M_\odot~{\rm yr}^{-1}$, where 
$M_\odot$ is the solar mass. 
Given that the wind needed to evacuate gas originally located around the progenitor, the expansion velocity of the cavity would have been slower than $v_{\rm w}$. 
Therefore, the above estimate of $\dot{M}$ should be treated as an upper limit. 
A calculation based on a more sophisticated model indeed leads to an estimate of $\dot{M} \sim 10^{-6}~M_\odot~{\rm yr}^{-1}$, which is 
within the range of typical value expected in an SD progenitor system \citep{zhou2016}. 
Thus, our discovery of shock deceleration strongly favors the SD origin of Tycho's SN.

The present work has demonstrated that expansion measurements of SNRs can serve as a powerful probe of their ambient gas environment, which 
contains key information about the progenitor systems. 
In the X-ray band, Chandra has observed SNRs multiple times with its superb angular resolution for 20 years and counting. 
X-ray observatories such as AXIS \citep{mushotzky2018} and Lynx \citep{gaskin2019}, with comparable angular resolution to Chandra, are currently planned for 2020s or beyond. 
Thus, we will be able to monitor the time evolution of expansion velocities of SNRs over a time span of $> 30~{\rm yr}$, and can 
expect more examples of witnessing ongoing deceleration of SNR expansion and further studies on SN progenitors. 
Specifically in the case of Tycho's SNR, such long-term monitoring would be important also to pin down the cause of the apparent inconsistency with expansion measurements in radio, 
which show much slower velocities for the past decades than what we observed before 2007. 
We finally note that we plan to perform another Chandra observation of the SNR next year, which will help us further confirm the present result and constrain more strongly the gas environment.

\acknowledgments
We thank the anonymous reviewers for their thoughtful comments.
This work is supported by JSPS Grant Nos. JP19H01936 (T.T.), JP19K03915 (H.U.), JP19H00704 (H.Y.), JP19K03913 (S.H.L.), JP18H05223 (K.M.), 
and JP20H00174 (K.M.). 
S.H.L. also acknowledges support from the Kyoto University Foundation and from the World Premier International Research Center Initiative (WPI), MEXT, Japan.



\end{document}